\documentclass[pra,twocolumn,nofootinbib, superscriptaddress]{revtex4}

\usepackage[dvipsnames]{xcolor}
\usepackage[utf8]{inputenc}
\usepackage{bm}
\usepackage{color}
\usepackage{hyperref}
\usepackage{amsmath, mathtools}
\usepackage{amssymb}
\usepackage{wasysym}
\usepackage{graphicx} 
\usepackage{booktabs} 
\usepackage{pdfsync}
\usepackage{verbatim}
\usepackage{latexsym}
\usepackage{dsfont}

\def\sec#1{\section{#1} }
\def\ssec#1{\subsection{#1} }

\def\({\left(}
\def\){\right)}
\def\[{\left[}
\def\]{\right]}

\def\a{\alpha}

\def\f#1#2{\frac{#1}{#2}}
\def\g{\gamma}

\def\d{\partial}
\def\de{\delta}

\def\del{\nabla}

\def\ep{\epsilon}

\def\k{\kappa}
\def\l{\lambda}
\def\L{\Lambda}

\def\o{\omega}

\def\p{\pi}
\def\r{\rho}

\def\th{\theta}

\def\ph{\phi}

\def\<{\langle}
\def\>{\rangle}

\newcommand{\ket}[1]{\left| #1 \right>} 
\newcommand{\bra}[1]{\left< #1 \right|} 

\providecommand{\abs}[1]{\left\lvert#1\right\rvert}

\usepackage{graphicx}
\usepackage{dcolumn}
\usepackage{bm}

\begin{document}

\preprint{APS/123-QED}

\title{Non-relativistic Effective Quantum Mechanics of the Coulomb Interaction}

\author{David M. Jacobs}\email{djacobs@norwich.edu}
\affiliation{Physics Department, Norwich University\\
158 Harmon Dr, Northfield, VT 05663}


\author{Matthew Jankowski}
\affiliation{Physics Department, Hamilton College\\
198 College Hill Rd., Clinton, NY 13323}

\date{\today}

\begin{abstract}
We apply the ideas of effective field theory to nonrelativistic quantum mechanics. Utilizing an artificial boundary of ignorance as a calculational tool, we develop the effective theory using boundary conditions to encode short-ranged effects that are deliberately not modeled; thus, the boundary conditions play a role similar to the effective action in field theory. Unitarity is temporarily violated in this method, but is preserved on average. As a demonstration of this approach, we consider the Coulomb interaction and find that this effective quantum mechanics can predict the bound state energies to very high accuracy with a small number of fitting parameters. It is also shown to be equivalent to the theory of quantum defects, but derived here using an \emph{effective} framework. The method respects electromagnetic gauge invariance and also can describe decays due to short-ranged interactions, such as those found in positronium. Effective quantum mechanics appears applicable for systems that admit analytic long-range descriptions, but whose short-ranged effects are not reliably or efficiently modeled. Potential applications of this approach include atomic and condensed matter systems, but it may also provide a useful perspective for the study of blackholes.
\end{abstract}

\pacs{Valid PACS appear here}
\maketitle

\sec{Introduction}\label{Sec:intro}

Effective field theory (EFT) has had many successes within subfields of physics that include condensed matter, particle physics, astrophysics, and cosmology.  The success of an effective field theory depends on a hierarchy of scales; the momenta or wavelengths of the experimental probes or observations of a system must be markedly different from the scale(s) of the processes not described, at least in detail, by the effective theory.

In this work we apply many of the ideas of EFT to quantum mechanics. The starting point of our discussion begins with the description of contact interactions, or delta-functions potentials, in quantum mechanics in two and three dimensions. It is known that such potentials sometimes require elaborate regularization and renormalization schemes to ensure physically sensible results are obtained  \cite{jackiw1995diverse}.  In  \cite{jackiw1995diverse} it was advocated that non-trivial boundary conditions are a preferable alternative method to using delta functions.

When applied to bound Coulomb states, this non-trivial boundary condition method -- also known as the method of self-adjoint extension -- has been shown to produce energy levels that obey Rydberg's formula, at least when the boundary condition parameter is small and proportional to the quantum defect \cite{BeckThesis}. In \cite{BeckThesis} it was shown that a unique boundary condition can provide an effective description of ``UV" physics near the origin, such as the effect of a finite nuclear radius or the Darwin fine-structure correction -- really anything that is, or may be approximated as a delta-function potential. There are two notable limitations to the analysis in \cite{BeckThesis}: (1) because all non-trivial $\ell\neq0$ solutions to the Schrodinger equation are not normalizeable, the method applies only for $s$-states ($\ell=0$) and (2) it does not reproduce the Rydberg-Ritz formula, the more accurate bound state energy formula in which the quantum defect is energy-dependent \cite{gallagher_1994}.

The motivation of \cite{Jacobs:2015han} was to extract a useful effective theory that would apply for \emph{all} angular momentum states.  In that work a finite region of space encompassing the origin was omitted from analysis, thereby naively obviating the need to discard the non-trivial $\ell\neq0$ solutions. The radius of what is referred to as the \emph{boundary of ignorance}, $r_b$ was interpreted as a kind of short-distance cutoff on which the boundary conditions effectively capture the omitted physics. In order to enforce unitarity conservation, however, the limit $r_b\to 0$ had to be taken at the end of any calculation; it follows that the analysis of \cite{Jacobs:2015han} only reproduces the results of \cite{BeckThesis}. Thereafter, results in \cite{Burgess:2016lal, Burgess:2016ddi} demonstrated similar results using an effective field theory of point particles.

In \cite{Jacobs:2019woc}, it was finally realized that violation of unitarity is not necessarily a problem if the violation is only temporary because the norm of a state function \emph{averages} to unity. Therefore it is not necessary to take the limit $r_b\to 0$ to arrive at a useful theory. From another point of view, one should not need to have a complete description of a system to arbitrarily short distance from the origin, just as the framework of EFT does not require a complete understanding of a theory to arbitrarily large momentum.

  Recall that for a one-dimensional system described by a scalar potential and a boundary at the coordinate $x=x_b$, all eigenmodes must obey the same (Robin) boundary condition in a standard analysis, i.e.
\begin{equation}
\psi_i(x_b)+{\cal Z} \psi_i'(x_b)=0\,,
\end{equation}
where the modes are labelled by generic index $i$, and ${\cal Z}$ is a real number; for example, ${\cal Z}=0$ corresponds to the Dirichlet condition.  The central equation of \cite{Jacobs:2019woc} is what results from promoting the boundary condition to be mode-dependent, i.e. ${\cal Z}\to {\cal Z}_i$, or
\begin{equation}\label{1d_EQM_bc}
\psi_i(x_b)+{\cal Z}_i \psi_i'(x_b)=0\,.
\end{equation}
Because the boundary condition varies for different eigenmodes, this theory is not \emph{instantaneously} hermitian or unitary; however, those standard conditions do hold when averaged over sufficiently long times\footnote{Unitarity violation appears to be a consequence of only considering the domain $x\geq x_b$, whereas a particle can in reality propagate in and out of the omitted region, $0\leq x <x_b$.}.   

In this article we significantly extend the analysis of \cite{Jacobs:2019woc}, demonstrating that this approach can be successfully applied in three dimensions with coupling to the full electromagnetic gauge field. We devote the bulk of our effort to analysis of hydrogenic atoms and arrive at the theory of quantum defects, albeit using a different framework from earlier analyses (see, e.g., a well-known review by Seaton \cite{Seaton_1983}).  The theoretical framing here is in the same spirit as that of effective field theory, however we do not appeal to a Lagrangian formalism. We start with the Schrodinger equation, using a Hamiltonian valid at long distance; the conditions of the wavefunction on the boundary of ignorance play a role analogous to the effective action. We therefore consider this method to be a demonstration of what can be called effective quantum mechanics.

In Section \ref{Gauge_invariance} we derive the three-dimensional version of the boundary condition \eqref{1d_EQM_bc} that respects electromagnetic gauge invariance and discuss its consequences. In Section \ref{BoundCoulombStates} we analyze the bound states of the Coulomb problem and derive the theory of quantum defects as a consequence of a low-energy effective theory describing the broken $SO(3)\otimes SO(3)$ symmetry of the Schrodinger-Coulomb problem. In Section \ref{Sec:EffectiveAndToy} we check the effective method against synthetically-generated data for a UV-complete model of an extended nucleus. In Section \ref{Sec:DataApplications} we consider the successes and limitations of this non-relativistic theory applied to real systems. In Section \ref{Sec:Decay} we consider decaying states. In Section \ref{Sec:Discussion} we conclude with a discussion of our results, list possible applications, and mention some outstanding issues. Throughout this article we use the natural unit convention $\hbar=c=1$.

\sec{Boundary Condition and Consequences}\label{Gauge_invariance}

The dynamics of a point charge of mass $m$ and charge $Q$ coupled electromagnetically is described by the Hamiltonian
\begin{equation}\label{Hamiltonian}
H=\f{\(\vec{p}-Q\vec{A}  \)^2}{2m}+Q\varphi\,.
\end{equation}
We expect that the boundary condition \eqref{1d_EQM_bc} can be promoted to a spherically symmetric and gauge-invariant version, namely
\begin{equation}\label{EQM_gaugeinvariant_bc}
R_i(r_\text{b}) + {\cal Z}_i(r_b)\, D_r R_i(r_\text{b})=0\,,
\end{equation}
where 
\begin{equation}\label{D_r_def}
D_r=\d_r - iQA_r\,,
\end{equation}
and $A_r$ is the radial component of the vector potential, $\vec{A}$. It would then follow that, under a local $U(1)$ transformation of the wave function and the electromagnetic field characterized by the function $\Omega$,
\begin{eqnarray}
\Psi &=& e^{i Q \Omega} \Psi' \notag\\
\vec{A}&=& \vec{A}'+\vec{\del} \Omega \notag\\
\varphi &=& \varphi'-\dot{\Omega} \,,
\end{eqnarray}
equation \eqref{EQM_gaugeinvariant_bc} will be invariant.

We derive \eqref{EQM_gaugeinvariant_bc} by mandating conservation of probability associated with \emph{a single eigenmode} where, for simpler notation, we note that $\Psi$ represents such a mode:
\begin{equation}\label{nondecaying_unitary_condition}
\f{d}{dt} (\Psi,\Psi) =-\int dV\,\del\cdot \vec{J} =0 \,,
\end{equation}
where the probability current density $\vec{J}$ following from the Hamiltonian \eqref{Hamiltonian} is
\begin{equation}\label{prob_curr_density}
\vec{J} = \f{i}{2m}\[\(\del \Psi^*\)\Psi- \Psi^* \(\del \Psi\) + 2iQ\vec{A} \abs{\Psi}^2\]\,.
\end{equation}
Given the spherically symmetric boundary at $r=r_b$, and separability of the eigenmode as
\begin{equation}\label{nondecaying_eigenmode}
\Psi=e^{-i\o t }R\(r\) Y_{\ell m}\(\th,\ph\)\,,
\end{equation}
the divergence theorem may be used to demonstrate that
\begin{equation}\label{bc_pre-equation}
\(D_r R\)^*R-R^* D_r R\bigg|_{r=r_b}=0\,,
\end{equation}
where $D_r$ is as defined in \eqref{D_r_def}. Following \cite{Bonneau:1999zq}, one may then multiply by an arbitrary constant $w$ with units of length and define the dimensionless complex quantities
\begin{eqnarray}\label{x_and_y_defs}
x&\equiv& R(r_b)\\
y&\equiv& wD_r R(r)|_{r=r_b}\,,
\end{eqnarray}
so that equation \eqref{bc_pre-equation} is then equivalent to 
\begin{equation}
\abs{x+iy}^2-\abs{x-iy}^2=0\,.
\end{equation}
The argument of the two terms above have an equal magnitude and differ only by an arbitrary phase factor $e^{i\th}$; it follows that
\begin{equation}
R(r_b)-\cot{\f{\th}{2}}w D_r R(r)\bigg|_{r=r_b}=0\,,
\end{equation}
which is boundary condition \eqref{EQM_gaugeinvariant_bc} once we make the identification
\begin{equation}\label{realZdef}
{\cal Z}=-\cot{\f{\th}{2}}w\,;
\end{equation}
again, we note that $\th$, $w$ and, therefore, ${\cal Z}$ are unique to the \emph{specific eigenmode} in question.

There are consequences of promoting the boundary condition to be mode-dependent. For example, the Hamiltonian is not Hermitian, which is observed by computing\footnote{The \emph{outward} normal to the boundary points in the inward radial direction, hence the change of sign in the last line of \eqref{Hamiltonian_diff}.}
\begin{eqnarray}\label{Hamiltonian_diff}
&&\<H\Psi_i,\Psi_j\> - \<\Psi_i,H\Psi_j\>\notag\\
&&=-\f{1}{2m}   \int dV\,\del\cdot \[\(\vec{D}\Psi_i\)^* \Psi_j - \Psi_i^*\(\vec{D}\Psi_j\)  \]\notag\\
&&=\f{1}{2m}r_b^2 \!  \int \!d\Omega  \[\(D_r\Psi_i\)^* \Psi_j - \Psi_i^*\(D_r\Psi_j\)  \]_{r=r_b},
\end{eqnarray}
which is not generally zero for two distinct eigenstates labeled by $i$ and $j$. In particular, because of the orthogonality of the spherical harmonics, this term is nonzero when states $i$ and $j$ have the same angular momentum quantum numbers\footnote{This would also include the spin quantum number if it were considered in this analysis.}.

If the angular quantum numbers are the same, e.g. $\ell_i=\ell_j$, etc., then
\begin{multline}\label{Hammy}
\<H\Psi_i,\Psi_j\> - \<\Psi_i,H\Psi_j\> \\
= \f{1}{2m}r_b^2 \({\cal Z}_i-{\cal Z}_j\)   \[\(D_rR_i\)^*\(D_rR_j\)  \]_{r=r_b} e^{-i\(\o_j-\o_i\)t}\,,
\end{multline}
 Equation \eqref{Hammy} is gauge invariant and never zero when $i\neq j$, but does average to zero over a period $2\pi/\abs{\o_i-\o_j}$; the same is true about the inner product of two distinct eigenmodes.  It may therefore be said that hermiticity and orthogonality do not generally hold at each instant, but they do in a time-averaged sense.
 
 Unitarity is also temporarily violated. By construction, the norm of each eigenmode is equal to unity for all time, but the same cannot be said for composite state. Following \cite{Jacobs:2019woc}, consider the state
\begin{equation}
\Upsilon = c_i \Psi_i + c_j \Psi_j\,,
\end{equation}
for two complex coefficients, $c_i$ and $c_j$. We assume the standard normalization condition
\begin{equation}\label{c_normalization condition}
\abs{c_1}^2 + \abs{c_2}^2 = 1\,,
\end{equation}
which will be justified below. The inner product of the composite state with itself is therefore
\begin{equation}
\< \Upsilon, \Upsilon\> = 1 + c_i^*c_j \<\Psi_i,\Psi_j \> + c_j^*c_i \<\Psi_j,\Psi_i \>\,.
\end{equation}
The last two offending terms do not vanish because the eigenmodes are not instantaneously orthogonal. Because the time derivative of the inner product is
\begin{equation}
\f{d}{dt}\<\Psi_i,\Psi_j \> = i\<H\Psi_i,\Psi_j\> + \text{cx. conj.}
\end{equation}
the time derivative of the composite state is
\begin{eqnarray}
\f{d}{dt}\< \Upsilon, \Upsilon\> &=& c_i^*c_j\f{d}{dt} \<\Psi_i,\Psi_j \> + \text{cx. conj.}\notag\\
&=&  -\({\cal Z}_i-{\cal Z}_j\)\abs{\r_{ij}} \sin{\[\(\o_i-\o_j\)t+ \th_{ij}\]}\,,
\end{eqnarray}
where
\begin{equation}
\r_{ij}=\f{c_i^*c_j}{m}r_b^2   \[\(D_rR_i\)^*\(D_rR_j\)  \] \bigg|_{r=r_b}
\end{equation}
and
\begin{equation}
\th_{ij}=\arg{\r_{ij}}\,.
\end{equation}
Apparently, the norm of the composite state is
\begin{equation}
\< \Upsilon, \Upsilon\> = 1 + \f{\({\cal Z}_i-{\cal Z}_j\)}{\o_i-\o_j} \abs{\r_{ij}} \cos{\[\(\o_i-\o_j\)t+ \th_{ij}\]}\,,
\end{equation}
which averages to unity over a period $2\pi/\abs{\o_i-\o_j}$.

The above analysis suggests that the best way to normalize a composite state is to demand
\begin{equation}
\< \Upsilon, \Upsilon\>_T \equiv 1\,,
\end{equation}
where the subscript indicates an averaging over a time, $T$ of appropriate length. A similar condition is apparently obeyed for the orthogonality of modes
\begin{equation}
\<\Psi_i,\Psi_j\>_T = 0\,,
\end{equation}
which justifies \eqref{c_normalization condition},  as well as the hermiticity of the Hamiltonian, 
\begin{eqnarray}
\<H\Psi_i,\Psi_j\>_T = \<\Psi_i,H\Psi_j\>_T \,.
\end{eqnarray}

Precisely what is considered \emph{appropriately} long depends on the physical system being studied. Because we focus almost exclusively on Coulomb bound states in the sections below, let us consider a composite state built from two such eigenmodes with the same angular quantum numbers. Clearly, averaging over a time longer than $2\pi/\abs{\o_i-\o_j}$ is sufficient according to the above analysis, and this is always  larger than the time scale associated with any omitted short-distance physics. $T$ must also be much shorter than any processes not included in this analysis that occur over relatively \emph{long} times, such as the spontaneous transition time between the two states, which is also true\footnote{
The transitions between states with the same angular momentum are accompanied by the emission of at least two photons. Such processes occur at a rate that does not exceed $\sim \abs{\o_i-\o_j}\a^5$ (see, e.g., \cite{Fitzpatrick_Quantum}).}. Although this by no means constitutes a proof that this method will work for all systems, it does suggest that its success as an effective theory depends on a clear hierarchy of \emph{time} scales, in addition to length scales.

\sec{Coulomb States}\label{BoundCoulombStates}

Here we consider an electron of charge $-e$ bound to a positive nucleus of charge $Ze$, so that the \emph{long-distance} Hamiltonian is given by equation \eqref{Hamiltonian} with $\vec{A}=0$ and scalar potential
\begin{equation}
\varphi = \f{Ze}{r}\,.
\end{equation}
The time-independent radial Schrodinger equation is
\begin{equation}
-\f{1}{r^2}\d_r \(r^2 R'(r)\) +\(\f{\ell\(\ell+1\)}{r^2}  -\f{2\k}{r} + q^2\)R(r) =0
\end{equation}
where 
\begin{equation}
\k=Zm\a\,,
\end{equation}
 $\a\simeq1/137$ is the fine structure constant,  and the energy eigenvalues are defined by
\begin{equation}\label{Energy_def}
E=-\f{q^2}{2m}\,.
\end{equation}
As described in \cite{Jacobs:2015han} there is one independent solution to this differential equation that is guaranteed to be square-integrable in the $r\to \infty$ limit\footnote{One of us (D.M.J.) would like to acknowledge Harsh Mathur for explaining the importance of this particular (decaying) linear combination of the two solutions to the confluent hypergeometric equation.}. We write this as, up to a normalization constant,
\begin{equation}\label{Exterior_Solution}
R(r) = e^{-qr} \(2qr\)^\ell \, U\(1+\ell -\f{\k}{q} \Big| 2(\ell+1) \Big| 2qr\)  \,,
\end{equation}
where $U$ is Tricomi's confluent hypergeometric function.

Quantization of the energies comes from application of the boundary condition \eqref{EQM_gaugeinvariant_bc}. It must be obeyed in such a way that any observables, such as the energy or, equivalently, $q$ are independent of the location of the boundary.  Because this is a long-distance effective theory, we expect the spatial scale of the wavefunction to be much larger than the boundary radius, or $qr_b\ll 1$. This means that, in principle, equation \eqref{EQM_gaugeinvariant_bc} could be expanded to arbitrary order in $qr_b$ and would then provide an arbitrarily precise analysis. Any function of $q$ could then be solved for; in particular, we find it best to solve for $\psi\(1+\ell -\f{\k}{q}\)$, where
\begin{equation}
 \psi(z) \equiv \f{\Gamma'(z)}{\Gamma(z)}
\end{equation}
is the digamma function. 

The digamma function is readily seen to appear in the series form of the Tricomi function, given in Appendix \ref{Appendix_Tricomi}. By using the digamma identity,
\begin{equation}\label{dg_identity}
\psi(1+z)=\psi(z)+\f{1}{z}\,,
\end{equation}
one could, in principle, solve equation \eqref{EQM_gaugeinvariant_bc} to write
\begin{equation}
\psi\(1+\ell -\f{\k}{q}\) = F_\ell\[{\cal Z}(r_b),r_b\]\,,
\end{equation}
where $F_\ell$ is a function of both the boundary function ${\cal Z}(r_b)$ and $r_b$, and accurate up to a particular order in the expansion parameter, $qr_b$. One could differentiate this equation with respect to $r_b$ and demand that it be equal to zero, resulting in a first order differential equation for the boundary function ${\cal Z}(r_b)$. It could then be said that ${\cal Z}(r_b)$ runs, in the sense of a renormalization group, with the boundary radius in way that ensures that the eigenvalues do not depend on where the boundary is; this was first described in \cite{Jacobs:2015han}, but a similar procedure may be found in \cite{Burgess:2016lal}.

However, at this point we simply integrate with respect to $r_b$ to implicitly solve for ${\cal Z}(r_b)$, i.e.
\begin{equation}
F_\ell\[{\cal Z}(r_b),r_b\] = \chi_\ell(q)\,,
\end{equation}
where $\chi_\ell(q)$ is an arbitrary integration function that must only be constant with respect to $r_b$. As in \cite{Jacobs:2019woc}, we posit that $\chi_\ell(q)$ captures the unspecified interactions behind the boundary, $r<r_b$. This establishes the result
\begin{equation}
\psi\(1+\ell -\f{\k}{q}\)=\chi_\ell(q)\,.
\end{equation}

We follow this apparently tautological procedure because, in practice, solving explicitly for ${\cal Z}(r_b)$ is cumbersome even at lowest order in the $qr_b$ expansion and for $\ell=0$; it becomes increasingly challenging at higher order and at higher values of $\ell$.  Instead, given the series form of the Tricomi function (see Appendix \ref{Appendix_Tricomi}) and boundary condition \eqref{EQM_gaugeinvariant_bc} that must be satisfied for arbitrary $r_b$, we make the generic ansatz\footnote{An arbitrary choice of length of $\(2\k\)^{-1}$ was put into the argument of the logarithm; any other choice can be made with a corresponding redefinition of the $c_j$'s. } for the boundary function,
\begin{equation}\label{Z_ansatz}
{\cal Z}(r_b) = \k^{-1}\sum_{j=1}^\infty \(c_j +d_j \ln{2\k r_b}   \) \(2\k r_b\)^j\,.
\end{equation}
By solving \eqref{EQM_gaugeinvariant_bc} for each term proportional to $r_b^j$ and $\ln{r_b}\,r_b^j$ we can determine the dimensionless coefficients $c_j$ and $d_j$ uniquely, up to the arbitrary integration function $\chi_\ell(q)$ which must appear in any equation containing $\psi\(1+\ell -\f{\k}{q}\)$. It may be verified that $\chi_\ell(q)$ only appears in the $c_j$ and does not first appear in the series until $c_{2\ell+2}$; it then appears in all subsequent $c_j$ which also may be understood through use of the digamma identity \eqref{dg_identity}.

Below we will not explicitly refer to $\chi_\ell(q)$, but its presence is implied in any discussion of $c_{2\ell+2}$, which we simply refer to as the integration function.

We do not, at present, have an analysis valid for arbitrary $\ell$. However, we have checked that the following procedure works at least up to $\ell=3$; it therefore seems implausible that it would not work to arbitrarily high $\ell$. We explicitly show the procedure for $\ell=0$ and $\ell=1$ below and the $\ell=2$ analysis may be found in Appendix \ref{Appendix_l_equal_2}. The summary is that at each $\ell$ we may write the solutions as deviations from their canonical form as
\begin{equation}\label{q_ansatz}
q=\f{\k}{n-\de_\ell}
\end{equation}
where $n$ is an integer and $\de_\ell$ is called the \emph{quantum defect} (see, e.g., \cite{Seaton_1983}). For each $\ell$-state we have considered it is possible to write the defect in the form
\begin{equation}\label{delta_expansion}
\de_\ell = \de_{\ell(0)}+\l_{\ell(1)} \f{E}{\L} + \l_{\ell(2)} \(\f{E}{\L}\)^2 + \dots\,,
\end{equation}
in other words, a low energy expansion in $E/\L$, where $\L$ is a high energy (UV) scale.

The connection between non-trivial boundary conditions and the quantum defect ansatz of \eqref{q_ansatz} was first made in \cite{BeckThesis}. The ansatz for $q$, equation \eqref{q_ansatz}, is a deviation from the canonical solutions,
\begin{equation}\label{q_canoncial}
q=\f{\k}{n}\,,
\end{equation}
and is motivated by two distinct considerations. The first reason is obvious: from an experimental point of view, hydrogenic atoms and highly-excited (Rydberg) states of large atoms are known to display spectra that are largely in agreement with \eqref{q_canoncial} -- this was, of course, one of the earliest successes of quantum mechanics. This canonical case apparently corresponds to the limit $\chi_\ell\to \pm\infty$, suggesting that it or, equivalently, the $c_{2\ell+2}$ will actually take on very large (but finite) values when this method is applied to real systems.

The second reason to use the canonical solutions as a point of departure is that they are special from the theoretical point of view; they are the unique solutions for which there is exists an $n^2$-fold  degeneracy at each energy level, $n$. This can be traced to presence of a ``hidden" $SO(3)\otimes SO(3)$ symmetry, the result of a conserved Runge-Lenz vector, in addition to angular momentum (see, e.g., \cite{weinberg2012lectures}).   As deviations are made from the canonical solution \eqref{q_canoncial}, therefore, one could say that the $SO(3)\otimes SO(3)$ symmetry is broken to the usual $SO(3)$ symmetry associated with 3-dimensional rotations \cite{BeckThesis}.   Although the Runge-Lenz vector operator continues to be conserved, the $\ell$-dependent boundary conditions mean that it acts on a different domain than that of the Hamiltonian, making it an unphysical operator\footnote{See, e.g.,  \cite{AlHashimi:2007ci} for a discussion about an analogous problem on a conical space in two dimensions.}.

\ssec{Effective description of $\ell=0$ bound states}\label{lzero_bound_states}

For the $\ell = 0$ solutions, the term-by-term consideration of the boundary condition \eqref{EQM_gaugeinvariant_bc} with the ansatz of \eqref{Z_ansatz} yields
\begin{equation}
c_2=-\gamma -\f{q}{4\k} - \f{1}{2}\[\ln{\f{q}{\k}} +\psi\(1-\f{\k}{q}\)\]\,,
\end{equation}
where $\g$ is the Euler-Mascheroni constant.  Performing an asymptotic expansion of the digamma function yields\footnote{Although this series does not converge, any truncation will be increasingly accurate as $\abs{E}$ decreases.}
\begin{equation}\label{c_2_eqn}
c_2=-\f{\p}{2} \cot{\p \f{\k}{q}}-\gamma -\f{1}{24} \f{E}{E_\text{Ry}} - \f{1}{240} \(\f{E}{E_\text{Ry}}\)^2 +{\cal O}\(\!\f{E}{E_\text{Ry}}\!\)^3   \,. 
\end{equation}
where
\begin{equation}
E_\text{Ry}\equiv\f{\k^2}{2m}\,.
\end{equation}

We make the ansatz 
\begin{equation}\label{q_defect_ansatz}
q=\f{\k}{n-\de}
\end{equation}
where, as argued in \cite{Jacobs:2019woc}, it should always be possible\footnote{In \cite{Jacobs:2019woc} the definition $n-\de\equiv\tilde{n}-\tilde{\de}$ was made, where $\tilde{n}$ is the closest integer to  $n-\de$, therefore $\abs{\tilde{\de}}<1/2$ was guaranteed. In the interest of clarity we do not adopt that notation; instead we mandate $\abs{\de}<1/2$.} to define $\abs{\de}\leq 1/2$. It appears that this is the only departure from the original quantum defect model, wherein there is no such restriction on the size of the defect (see, e.g., \cite{gallagher_1994}).  We expand equation \eqref{c_2_eqn} in small $\de$ and find that the defect can be solved for implicitly as
\begin{multline}\label{shitballs_0}
\de =     \f{\(1-\f{\p^2}{3}\de^2 -\f{\p^4}{45}\de^4  + {\cal O}\(\de^6\)\)}{2c_2 + 2\gamma +\f{1}{12} \f{E}{E_\text{Ry}} + \f{1}{120} \(\f{E}{E_\text{Ry}}\)^2 +{\cal O}\(\f{E}{E_\text{Ry}}\)^3 } \,. 
\end{multline}

We do not know what functional form the integration function -- or $c_2$ -- should have, but two comments are warranted. Firstly, deviations from the canonical Coulomb spectrum are assumed here to be the result of short-ranged/high-energy physics \emph{not included explicitly} in the Coulomb potential, and therefore we expect those deviations not to depend explicitly on the ratio $E/E_\text{Ry}$. Secondly, a series form for $c_2$ as an expansion in $E$ over some high energy scale is arguably the simplest guess, and is also consistent with the well-known and successful approach taken when writing down an effective action in the context of effective field theory. Not knowing \emph{a priori} what the coefficients of this expansion should be, we parametrize the denominator of equation \eqref{shitballs_0} to be in the series form
\begin{equation}
A_0 + A_1 \f{E}{\L} + A_2 \(\f{E}{\L}\)^2 +\dots\,,
\end{equation}
where $\L$ is a high energy scale and, in the parlance of field theory, we call the $A_i$  \emph{renormalized} expansion coefficients. Equivalently, the integration function could apparently be written
\begin{equation}
c_2 =B_0 + B_1\f{E}{\L}+ B_2\f{E^2}{\L^2} +\dots\,,
\end{equation}
where the \emph{bare} expansion coefficients, $B_i$ are related to their renormalized counterparts by
\begin{eqnarray}
B_0&=&\f{A_0}{2}-\gamma\notag\\
B_1&=&\f{A_1}{2}  - \f{1}{24} \f{\L}{E_\text{Ry}}\notag\\
B_2&=&\f{A_2}{2} -\f{1}{240} \f{\L^2}{E_\text{Ry}^2}\,,
\end{eqnarray}
and so on. Summarizing, we have
\begin{equation}\label{delta_summary}
\de =     \f{\(1-\f{\p^2}{3}\de^2 -\f{\p^4}{45}\de^4  + {\cal O}\(\de^6\)\)}{A_0 + A_1 \f{E}{\L} + A_2 \(\f{E}{\L}\)^2 +\dots} \,, 
\end{equation}
which can be iteratively solved for $\de$. Without any loss of generality, we therefore write
\begin{equation}\label{generic_delta}
\de = \de_{0}+\l_{1} \f{E}{\L} + \l_{2} \(\f{E}{\L}\)^2 + \dots\,,
\end{equation}
where $\de_0$ and the $\l_i$ are dimensionless coefficients.  One could speculate that, because $\de\to0$ as the canonical solutions are recovered, $\de_0$ is proportional to  $E_\text{Ry}/\L$, possibly raised to a positive power. We demonstrate below that this is indeed the case, at least when applied to the hydrogen atom.

\ssec{Effective description of $\ell=1$ bound states}

Following the procedure used in the previous section, for $\ell=1$ we discover
\begin{multline}
c_4=\f{9-16\g}{256} -\f{3}{64}\f{q}{\k} -\f{\(1-\g\)}{16}\f{q^2}{\k^2} + \f{1}{64}\f{q^3}{\k^3}\\
+\f{q^2-\k^2}{32\k^2}\[\ln{\f{q}{\k}}+\psi\(2-\f{\k}{q}\)\]\,.
\end{multline}
After expanding the digamma function in small $q/\k$ and writing this expression in terms of energies we find
\begin{equation}\label{shitballs_1}
\de =\f{\(1+\f{E}{E_\text{Ry}}\)\(1-\f{\p^2}{3}\de^2 -\f{\p^4}{45}\de^4  + {\cal O}\(\de^6\)\)}{32c_4 \!-\! \f{9}{8} \!+\! 2\g \!-\! \f{11-24\g}{12}\f{E}{E_\text{Ry}} \!+\! \f{11}{120} \(\f{E}{E_\text{Ry}}\)^2 \!+\! {\cal O}\(\f{E}{E_\text{Ry}}\)^3}
   \,.
\end{equation}
Here a convenient parametrization of the denominator in equation \eqref{shitballs_1} is the series form
\begin{equation}
\(1+\f{E}{E_\text{Ry}}\)\(A_0 + A_1 \f{E}{\L} + A_2 \(\f{E}{\L}\)^2 +\dots\)\,,
\end{equation}
or, equivalently,
\begin{equation}
c_4 =B_0 + B_1\f{E}{\L}+ B_2\f{E^2}{\L^2} +\dots
\end{equation}
where
\begin{eqnarray}
B_0&=&\f{A_0}{32} + \f{9}{256}-\f{\gamma}{16}\notag\\
B_1&=&\f{A_1}{32} + \(\f{11}{384} - \f{\g}{16}\) \f{\L}{E_\text{Ry}}  \notag\\
B_2&=&\f{A_2}{32} - \f{11}{3840} \(\f{\L}{E_\text{Ry}}\)^2\,.
\end{eqnarray}
The result is the same as that of the $\ell=0$ case, namely $\de$ may be put in a form identical to equations \eqref{delta_summary} and \eqref{generic_delta}. The $\ell=2$ analysis follows similarly and may be found in Appendix \ref{Appendix_l_equal_2}.

\ssec{Brief comments about scattering}

Because of the apparent equivalence between this effective approach and that of quantum defect theory we do not dwell on the analysis of scattering states. We simply note that, whereas in bound state calculations the definition $q^2=-2mE$ is made, for scattering one defines the wave number $k$ by
\begin{equation}
k^2=2mE\,,
\end{equation}
where the energy, $E>0$. This suggests an analytic continuation of the integration function $\chi_\ell(q^2)$ in the variable to $q^2\to -k^2$, in other words an analytic continuation of the defects $\de_\ell(E)$ from $E<0$ to $E>0$. This is precisely what is known to occur within quantum defect theory and we direct the interested reader toward the relevant literature (see, e.g., \cite{Seaton_1983} and references therein).

\sec{Fits to synthetic data}\label{Sec:EffectiveAndToy}

Here we consider the long-range Coulomb potential modified at short distance with a specific UV-completion, namely one in which there is a constant  ``nuclear" charge density. The scalar potential is therefore
\begin{equation}
\varphi= 
\begin{cases}
\f{Zer^2}{R_\text{nuc}^3},~~~~~&(0\leq r \leq R_\text{nuc})\\
\f{Ze}{r},~~~~~~~&(r>R_\text{nuc})\,.
\end{cases}
\end{equation}

The time-independent radial Schrodinger equation in the nuclear interior is
\begin{equation}\label{Interior_Schrodinger_eqn}
-\f{1}{r^2}\d_r \(r^2 R'(r)\) +\(\f{\ell\(\ell+1\)}{r^2}  -\f{4r^2}{b^4} + q^2\)R(r) =0
\end{equation}
where 
\begin{equation}
b^4\equiv \f{2R_\text{nuc}^3}{Zm\a}\,.
\end{equation}
Imposing regularity at the origin, the solution to \eqref{Interior_Schrodinger_eqn} may be written, up to a normalization constant, as
\begin{multline}\label{Interior_Solution}
R(r) = e^{-i\(\f{r}{b}\)^2} \(\f{r}{b}\)^\ell\\
\times M\(\f{2\ell+3}{4} -\f{i}{8}\(qb\)^2 \Big| \f{2\ell+3}{2} \Big| 2i\(\f{r}{b}\)^2\)  \,,
\end{multline}
where $M$ is Kummer's hypergeometric function.

It is important to separately consider two different types of hydrogenic systems, namely those in which deviations from the Coulomb potential occur at radii that are smaller and larger than the Bohr radius. For this reason we consider two examples in which the nuclear radius, $R_\text{nuc}$ satisfies either or $\k R_\text{nuc}<1$ or $\k R_\text{nuc}>1$. The synthetic bound state energies of this UV-complete model are generated by matching the interior and exterior solutions, equations \eqref{Exterior_Solution} and \eqref{Interior_Solution}, and their first derivatives at $r=R_\text{nuc}$.

To display the robustness of the effective theory we apply it to bound states with a leading order (LO) fit using only $\de_{\ell(0)}$, next-to-leading order (NLO) by fitting for $\de_{\ell(0)}$ and $\l_{\ell(1)}$, and next-to-next-to leading order (NNLO) by fitting for $\de_{\ell(0)}$, $\l_{\ell(1)}$, and $\l_{\ell(2)}$. We assume both $m$ and $\a$ are perfectly known by some independent means, and utilize equations \eqref{Energy_def}, \eqref{q_ansatz}, and \eqref{delta_expansion} to fit to the lowest energy levels, i.e. those with the largest $\abs{E}$, so as to make predictions for the higher energy levels.

\begin{figure}[htp]
  \begin{center}
    \includegraphics[scale=.75]{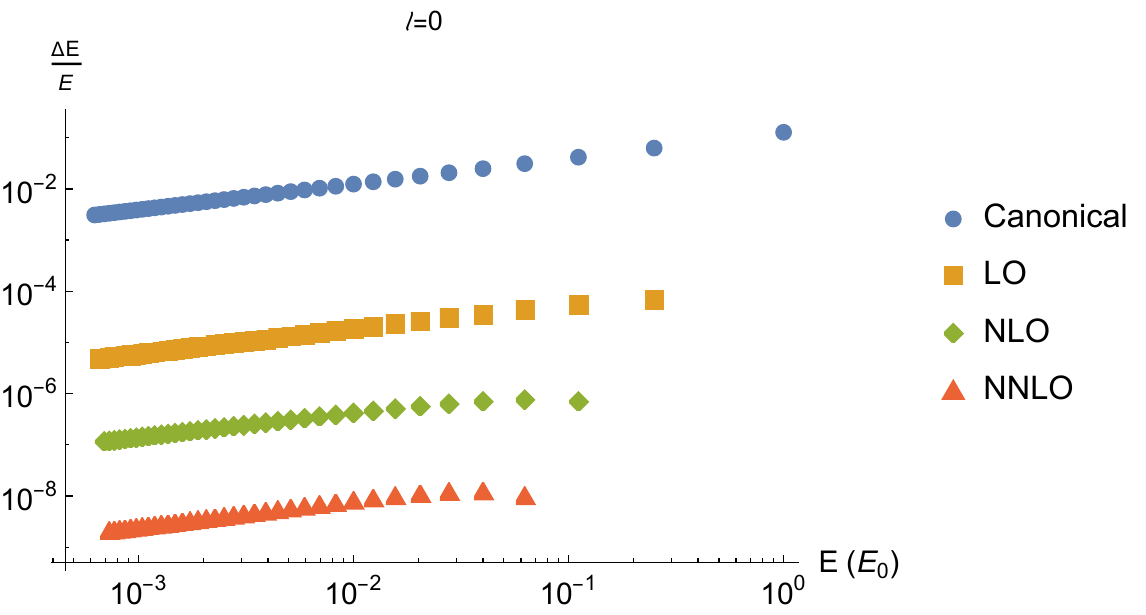}
  \end{center}
  \caption{Relative errors in the $\ell=0$ binding energies, compared to the UV-complete model, wherein $\k R_\text{nuc}=0.31$. }
  \label{pt31_l_0}
  \end{figure}

\begin{figure}[htp]
  \begin{center}
    \includegraphics[scale=.75]{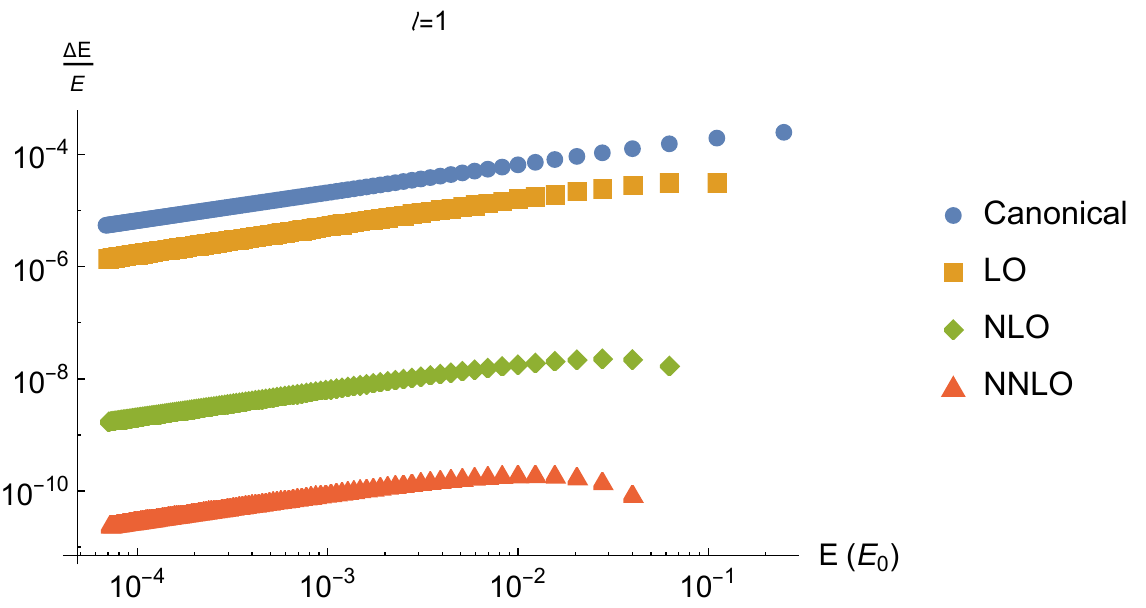}
  \end{center}
  \caption{Relative errors in the $\ell=1$ binding energies, compared to the UV-complete model wherein $\k R_\text{nuc}=0.31$, as in Figure \ref{pt31_l_0}.}
  \label{pt31_l_1}
  \end{figure}

In Figures \ref{pt31_l_0} and \ref{pt31_l_1} we display the relative error in the predicted energy levels for $\k R_\text{nuc}=0.31$ when $\ell=0$ and $\ell=1$, respectively; in those figures we normalize the energies to the ground state, $E_0$.

\begin{figure}[htp]
  \begin{center}
    \includegraphics[scale=.75]{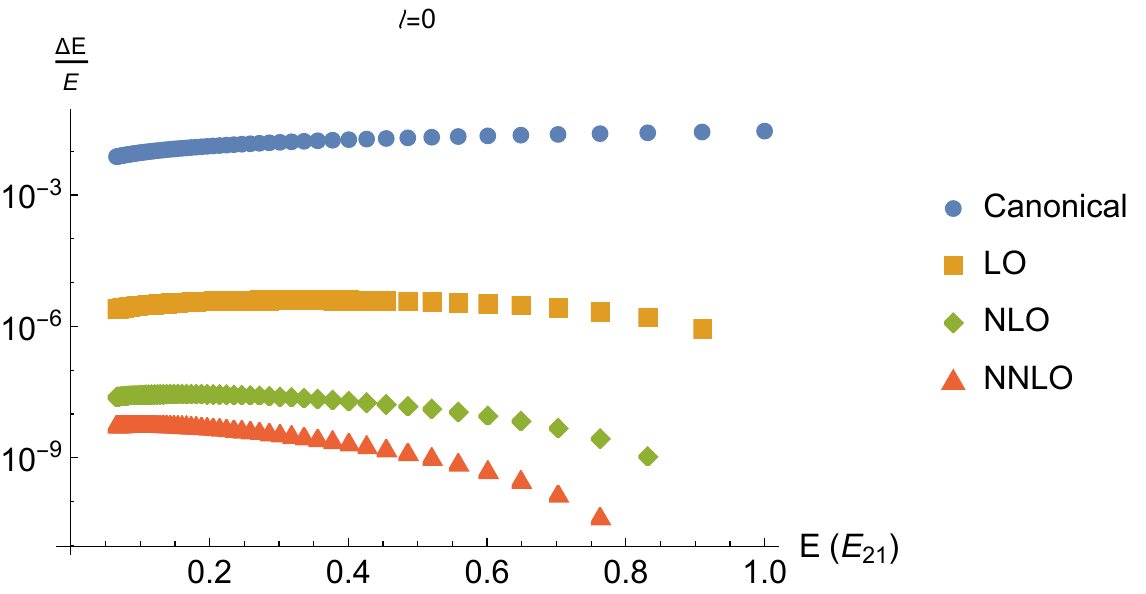}
  \end{center}
  \caption{Relative errors in the $\ell=0$ binding energies, compared to the UV-complete model, wherein $\k R_\text{nuc}=2.73$.}
  \label{2pt73_l_0}
  \end{figure}

\begin{figure}[htp]
  \begin{center}
    \includegraphics[scale=.75]{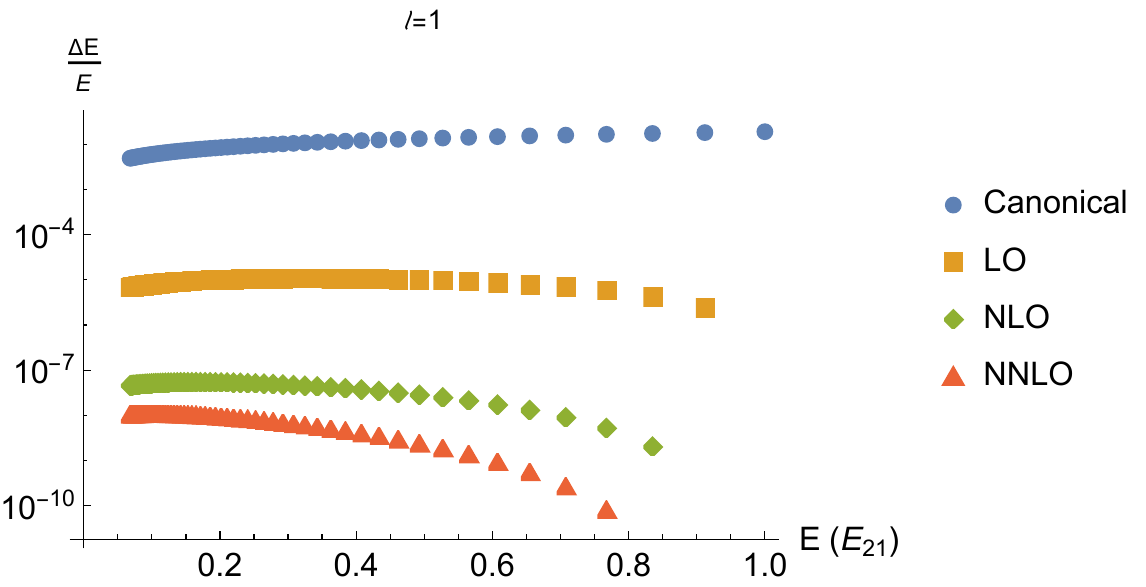}
  \end{center}
  \caption{Relative errors in the $\ell=1$ binding energies, compared to the UV-complete model, wherein $\k R_\text{nuc}=2.73$, as in Figure \ref{2pt73_l_0}.}
  \label{2pt73_l_1}
  \end{figure}

 For the large nuclear radius we choose $\k R_\text{nuc}=2.73$. The method here is unsuccessful unless one fits the effective theory to the synthetic data starting at somewhat higher energy levels. This is consistent with the fact that this method is a \emph{long-distance} effective theory; thus one should only expect it to provide accurate predictions when the characteristic length scale of the wavefunction is large compared to the nuclear radius, or $q R_\text{nuc}<1$. We choose not to fit the first 20 synthetic levels (ground state and 19 excited states), beginning our fits near the canonical eigenvalue of $q=\k/21$ and can therefore make predictions beginning near $q=\k/22$. In Figures \ref{2pt73_l_0} and \ref{2pt73_l_1} we display our results for $\ell=0$ and $\ell=1$, respectively, normalizing the energies to the 21st excited state, $E_{21}$. Although the errors initially grow marginally as higher energy levels are considered, eventually there is a turnover and the errors begin to decrease. In any case, at any given energy level,  the effective method gives predictions that are always more accurate at higher order.

\sec{Application to physical systems}\label{Sec:DataApplications}

In so far as Rydberg atoms are concerned, the $\de$ ansatz of \eqref{delta_expansion} is equivalent to the usual quantum defect method(s), wherein the modified Rydberg-Ritz expression is often written 
\begin{equation}\label{Gallagher_defect}
\de_{n\ell j} = \de_0 + \f{\de_2}{\(n-\de_0\)^2} + \f{\de_4}{\(n-\de_0\)^4} + \dots\,,
\end{equation}
for some experimentally determined constants $\de_0, \de_2$, etc. \cite{gallagher_1994}. The only difference with our approach, as mentioned in Section \ref{lzero_bound_states}, is that here we restrict the size of the defect to obey $\abs{\de_0}<1$. For example, the measured transition frequencies of the alkalis $^{23}\text{Na}$, $^{39}\text{K}$, and $^{85}\text{Rb}$ are fit with the original defect model to give a leading order $s$-state defect, $\de_{0}^\text{QDT}\simeq1.380$, $2.180$ and $3.131$, respectively \cite{gallagher_1994}. In the effective model these are simply interpreted as $\de_0=0.380$, $0.180$, and $0.131$. In other words, the ground state of an alkali atom corresponds to $n=1$, rather than the principal quantum number corresponding to its row in the periodic table.

Fitting the hydrogen spectrum, for example, one can achieve reasonably accurate results; however, the predictions become only marginally more accurate at higher order in the effective theory, and this is likely because of relativistic effects that are not accounted for. Consider the effective theory applied to a particular state of a hydrogenic atom in which the nucleus has a charge $+e$. At leading order, $\de_\ell = \de_{\ell(0)}$ so that, expanding in small $\de_{\ell(0)}$, the energy levels are
\begin{equation}\label{lowest_order_hydrogen}
E=-\f{m\a^2}{2n^2} -\f{m\a^2}{n^3}\de_{\ell(0)} 
 -\f{3m\a^2}{2n^4}\de_{\ell(0)}^2  + \dots
\end{equation}
The first term corresponds to the canonical eigenvalues, whereas the second term is proportional to the corrections that are usually obtained using perturbation theory; in particular, short-ranged corrections to the Coulomb potential proportional to $1/r^2$, $1/r^3$, etc., as well as a delta-function centered about $r=0$ give corrections proportional to $n^{-3}$  (see, e.g., \cite{PhysRevA.42.1123}). Let us call those potential corrections $U_\text{rel}$.  Perturbation theory is used to correct the canonical energy levels by an amount 
\begin{equation}\label{Perturbative_energy_correction}
\bra{\Psi} U_\text{rel}\ket{\Psi} = - \f{m\a^4}{\(2j+1\)n^3}\,, 
\end{equation}
where $j=\ell\pm1/2$ \cite{berestetskii1982quantum}. Matching with the effective method indicates that
\begin{equation}
\de_{\ell(0)} = {\cal O}(\a^2)\,,
\end{equation}
or $\de_{\ell(0)} \propto E_\text{Ry}/\L$, which is true when we set the high energy scale $\L=m$, the mass of the electron.

At the next-to-leading order, we apparently have
\begin{equation}
\de_\ell = \de_{\ell(0)}-\l_{\ell(1)}\f{\a^2}{2n^2}\,.
\end{equation}
With an additional parameter there is, of course, an improved fit to the hydrogen spectrum; however, it is only a marginal improvement. Although equation \eqref{lowest_order_hydrogen} is modified, the effect of the parameter $\l_{\ell(1)}$ only appears at order $n^{-5}$, whereas there is already trouble with the order $n^{-4}$ term. This is because there is a remaining fine-structure effect not captured by $U_\text{rel}$, but instead comes from the relativistic correction to the kinetic energy. That kinetic correction amounts to
\begin{equation}
+\f{3m\a^4}{8n^4}
\end{equation}
for all $\ell$ states of Hydrogen, which cannot be accounted for simultaneously with the order $n^{-3}$ correction in equation \eqref{Perturbative_energy_correction}. In any case it would not be appropriate;  the relativistic correction to the kinetic energy is not a short-ranged effect that should be hidden behind the boundary of ignorance.

Although this and the preceding sections demonstrate the utility of the non-relativistic effective quantum mechanics, a relativistically corrected version of the theory is clearly warranted. Those results will appear in forthcoming work  \cite{JacobsPs}.

\sec{Decays due to UV effects}\label{Sec:Decay}

Here we consider if and how the analysis must be modified if the eigenmode in question decays at a rate, $\Gamma$, via some interaction(s) near the origin\footnote{This effective method can describe decays of states due to short-ranged effects, such as annihilation; the analysis presented here does not describe, e.g., transitions from one state to another.}. In other words, normalizing the state at $t=0$,
\begin{equation}
\<\Psi,\Psi\>=e^{-\Gamma t}\,,
\end{equation}
where, for simpler notation, we note that $\Psi$ represents a single time-dependent eigenmode. We assume the eigenmodes may be written in the variable-separated form as
\begin{equation}\label{decaying_eigenmode}
\Psi=e^{-i\o t - \f{\Gamma}{2}t  }  R\(r\) Y_{\ell m}\(\th,\ph\)\,,
\end{equation}
where $\o$ is real and any normalization constant is absorbed into $R(r)$. In the case of a decaying state we must modify equation \eqref{nondecaying_unitary_condition} to 
\begin{equation}\label{decaying_unitary_condition}
\f{d}{dt} \<\Psi,\Psi\> =-\int dV\,\del\cdot \vec{J} =-\Gamma e^{-\Gamma t} \,.
\end{equation}
The probability current density is still given by equation \eqref{prob_curr_density}, but here the application of the divergence theorem results in a modification to equation \eqref{bc_pre-equation}, namely
\begin{equation}\label{bc_pre-equation_w_decay}
\(D_r R\)^*R-R^* D_r R\bigg|_{r=r_b}=\f{2im}{r_b^2}\Gamma\,.
\end{equation}
Multiplying this equation by $w$, an arbitrary constant with units of length and making the same definitions for $x$ and $y$ as in equation \eqref{x_and_y_defs}, it may be verified that equation \eqref{bc_pre-equation_w_decay} is equivalent to
\begin{equation}
\abs{x+iy\(1-\f{mw\Gamma}{r_b^2\abs{y}^2}\)}^2=\abs{x-iy\(1+\f{mw\Gamma}{r_b^2\abs{y}^2}\)}^2\,.
\end{equation}
Because there is an equivalence of the arguments up to a phase factor $e^{i\th}$, one may write
\begin{equation}
R-\(\cot{\f{\th}{2}}w +i\f{m\Gamma}{r_b^2\abs{D_r R}^2} \)D_r R=0\,.
\end{equation}
It follows that the boundary condition is a complexified form of \eqref{EQM_gaugeinvariant_bc}, namely
\begin{equation}\label{EQM_gaugeinvariant_bc_w_decay}
R(r_b)+\({\cal Z}_\text{re}(r_b)+i{\cal Z}_\text{im}(r_b)\)D_r R(r_b)=0\,,
\end{equation}
where
\begin{equation}
{\cal Z}_\text{re}(r_b)=-\cot{\f{\th}{2}}w\,,
\end{equation}
and
\begin{equation}\label{ImZ_eqn}
{\cal Z}_\text{im}(r_b)=-\f{m\Gamma}{\abs{r_b D_r R}^2}\,.
\end{equation}

 The presence of $D_r R\(r_b\)$ in the definition ${\cal Z}_\text{im}(r_b)$ may seem strange because it suggests a non-linearly realized boundary condition; however, we remind the reader that this analysis is meant to be approximate. We have shown that if the eigenmode is \emph{exactly} proportional to $e^{-\f{\Gamma}{2} t}$, then it follows that equation \eqref{ImZ_eqn} must hold \emph{exactly}. However, such a rigid expectation is inappropriate. Clearly, this method is not capable of describing, e.g., deviations from a pure exponential decay law which is known to occur at very short times after an unstable state has been established (see, e.g., \cite{Chiu:1977ds}). The method described here is intended as a long-distance -- hence a long-time --  effective theory. We therefore suggest that equation \eqref{ImZ_eqn} gives only a qualitative relationship: ${\cal Z}_\text{im}(r_b)$ is both proportional to the decay rate of the system and bounded according to ${\cal Z}_\text{im}(r_b)\leq0$. Beyond these qualitative features, we only assume that ${\cal Z}_\text{im}(r_b)$ has \emph{some} dependence on $r_b$ that can be determined in the analysis of a particular system.

Consider positronium, a system described at long distance by a Coulomb potential with a reduced mass of $m/2$. The analysis from Section \ref{BoundCoulombStates} follows in a nearly identical fashion, but the energy eigenvalues and quantum defects are complex, i.e.
\begin{equation}\label{cx_E}
E=-\f{q^2}{m}=\o  -i \f{\Gamma}{2}\,,
\end{equation}
where
\begin{equation}\label{Ps_q_ansatz}
q=\f{m\a}{2\(n-\de\)}\,,
\end{equation}
and, at lowest order,
\begin{equation}\label{cx_delta_0}
    \de = \de_{0,\text{re}} + i  \de_{0,\text{im}}\,.
\end{equation}

We will further assume that $\de_{0,\text{im}}\ll1$ so that a perturbative expansion in small $\de$ is still possible and therefore the analysis of Section \ref{BoundCoulombStates} is equally valid; we confirm this below. From equations \eqref{cx_E}, \eqref{Ps_q_ansatz}, and \eqref{cx_delta_0}, it is apparent that the real part of the energy
\begin{equation}
\o = -\f{m\(Z\a\)^2}{4n^2} -\f{m\(Z\a\)^2}{2n^3}\de_{0,\text{re}} +\dots\,,
\end{equation}
whereas the decay rate is given by
\begin{equation}\label{EQM_Decay_rate}
\Gamma = \f{m\(Z\a\)^2}{n^3}\de_{0,\text{im}} +\dots\,,
\end{equation}
which displays the standard $n^{-3}$ dependence expected from equation \eqref{Perturbative_energy_correction}. At higher order, $\de_{0,\text{im}}$ would affect $\o$ as well; however, we have already established in Section \ref{Sec:DataApplications} that this analysis is limited because it is missing relativistic corrections and therefore only these lowest-order results are worth reporting here.

Within Quantum Electrodynamics, the lowest order decay rate of positronium is predicted to be (see, e.g., \cite{Cassidy:2018}) 
\begin{equation}
\Gamma_\text{QED}=
\begin{cases}
\f{m\a^5}{2n^3}~~~~~~&\text{(singlet)}\\
\f{4}{9\pi}\(\pi^2-9\)\f{m\a^6}{2n^3}~~~~~~&\text{(triplet)}
\end{cases}
\end{equation}
which means that matching to that UV-complete theory would yield
\begin{equation}
\de_{0,\text{im}}=
\begin{cases}
{\cal O}(\a^3)~~~~~~&\text{(singlet)}\\
{\cal O}(\a^4)~~~~~~&\text{(triplet)}\,.
\end{cases}
\end{equation}

\sec{Discussion}\label{Sec:Discussion}

We have shown how to construct a nonrelativistic effective quantum mechanics in three dimensions for systems possessing spherical symmetry.  The short-distance cutoff length, $r_b$ is a conceptual and calculational crutch used to derive our results and ultimately vanishes from any final result. The role of the boundary function is that of a coupling constant; each mode ``feels" a different coupling constant that varies with energy. A high energy scale, $\Lambda$ appears in the low-energy expansion of physical quantities, such as bound states. We focused primarily on the Coulomb interaction and have found non-trivial results for \emph{all} angular momentum states, ultimately showing an equivalence to quantum defect theory. We have also shown the method provide a means of describing decays due to effects at short distance.

The most pressing question is how to apply this approach relativistically for application to high precision spectroscopy of atoms and molecules. Rydberg atoms, are of particular importance because they have potential applications in quantum computing and electromagnetic field sensing, for example \cite{Adams_2019}.  There also appears to be a pertinent application to positronium, in particular because of a recently discovered discrepancy between a measured transition frequency in that system and the predictions from QED \cite{Gurung:2020hms}. These ideas presumably also have applications in the areas of condensed matter, particle physics, and possibly gravitation. Like the Coulomb interaction, blackholes provide a $1/r$ potential at long distances and exhibit a kind of boundary, the event horizon, behind which information is obscured. 
\\\\\\\\
\begin{center}
{\bf Acknowledgements}\\
\end{center}
Thanks are owed to Harsh Mathur, with whom many discussions were had during early stages of this work. Thanks are also owed to the late Bryan Lynn for raising the question about decaying systems and positronium, in particular. One of us (MJ) would like to thank Hamilton College for funding during a portion of this work and another of us (DMJ) would like to thank the Hamilton College Physics Department for its hospitality  during early stages of this work.

\appendix

\sec{Series form of the Tricomi function}\label{Appendix_Tricomi}
The solutions to the Schrodinger-Coulomb problem involve solutions to the confluent hypergeometric equation
\begin{equation}\label{CHE}
y g''(y) + \(B-y\)g'(y)    -Ag(y)=0\,.
\end{equation}
A standard textbook analysis, e.g., \cite{arfken2005mathematical}, involves a series ansatz one may show that gives two independent solutions
\begin{equation}
M(A,B,y)=\sum_{n=0}^\infty  \f{1}{n!}\f{A^{(n)}}{B^{(n)}} y^{n}\,,
\end{equation}
known as Kummer's function, and
\begin{equation}
M_2(A,B,y)=y^{1-B}  M(A+1-B,2-B,y)
\end{equation}
The Tricomi function, $U(A,B,y)$ is the special linear combination of the two that is guaranteed to decay as $y\to \infty$, usually defined as
\begin{multline}
U(A,B,y)\equiv  \f{\Gamma{(1-B)}}{\Gamma{(A-B+1)}}  M(A,B,y)\\
+ \f{\Gamma(B-1)}{\Gamma(A)} M_2(A,B,y)\,;
\end{multline}
this along with, e.g., $M(A,B,y)$ may be chosen as a linearly independent set of solutions as long as $B$ is not an integer greater than $1$.

In the Coulomb problem, however, $B=2+2\ell$, so care must be taken to understand the series form of $U(A,B,y)$. One can, for example, let $B=2+2\ell +\ep$, where $\ep$ is treated as perturbatively small; in the end one can let $\ep\to0$ and show that the Tricomi function may be written exactly as
\begin{multline}
U\(A,2+2\ell,y\) = \f{1}{\Gamma\(A\)\Gamma\(A-2\ell-1\)}\times\\
\(-\sum_{n=0}^{2\ell}\f{\(-y\)^{n-2\ell-1}}{n!}\Gamma\(A-2\ell-1+n\)\Gamma\(2\ell+1-n\)   \right.\\
\left. + \sum_{n=0}^{\infty}\f{y^n}{n!}\f{\Gamma(A+n)}{\Gamma(2\ell+2+n)}\biggl[\psi(A+n)-\psi(2\ell+2+n)\right.\\
-\psi(n+1) +\ln{y}   \biggr]   \Biggr) 
\end{multline}

\sec{Effective description of $\ell=2$ bound states}\label{Appendix_l_equal_2}

Here we find
\begin{multline}
c_6=\f{59-72\g}{373\,248} -\f{5}{20\,736}\f{q}{\k}  + \f{\(180\g-197\)}{186\,624}\f{q^2}{\k^2}\\
 + \f{5}{6912}\f{q^3}{\k^3}   + \f{\(35-24\g\)}{31\,104}\f{q^4}{\k^4}  - \f{1}{5184}\f{q^5}{\k^5}\\
-\f{\(4q^4-5q^2\k^2+\k^4\)}{10368\k^4}\[\ln{\f{q}{\k}}+\psi\(3-\f{\k}{q}\)\]\,.
\end{multline}
After expanding the digamma function in small $q/\k$ and writing this in terms of energies we find
\begin{multline}
c_6= -\f{\p}{10368}\(1+\f{5E}{E_\text{Ry}}+\f{4E^2}{E_\text{Ry}^2}\)\cot{\p\f{\k}{q}}\\
 + \f{59-72\g}{373\,248} - \f{360\g-283}{373\,248} \(\f{E}{E_\text{Ry}}\)\\
  -\f{629-960\g}{1224160}\(\f{E}{E_\text{Ry}}\)^2  +{\cal O}\(\f{E}{E_\text{Ry}}\)^3\,.
\end{multline}
We make the defect ansatz in equation \eqref{q_defect_ansatz} to find
\begin{multline}\label{shitballs_2}
\de =\(1+\f{5E}{E_\text{Ry}}+\f{4E^2}{E_\text{Ry}^2}\)\(1-\f{\p^2}{3}\de^2 -\f{\p^4}{45}\de^4  + {\cal O}\(\de^6\)\)\\
\times\[
10368c_6 - \f{59}{36} + 2\g - \(\f{283}{36}-10\g\)\f{E}{E_\text{Ry}}\right.\\
\left. + \(\f{629}{120} -8\g\)\(\f{E}{E_\text{Ry}}\)^2 + {\cal O}\(\f{E}{E_\text{Ry}}\)^3  \]^{-1} \,.
\end{multline}
We may parametrize the denominator of equation \eqref{shitballs_2} to be in the series form
\begin{equation}
\(1+\f{5E}{E_\text{Ry}}+\f{4E^2}{E_\text{Ry}^2}\)\(A_0 + A_1 \f{E}{\L} + A_2 \(\f{E}{\L}\)^2 +\dots\)\,,
\end{equation}
or, equivalently,
\begin{equation}
c_6 =B_0 + B_1\f{E}{\L}+ B_2\f{E^2}{\L^2} +\dots
\end{equation}
where
\begin{eqnarray}
B_0&=&\f{A_0}{10\,368} + \f{59}{373\,248}-\f{\gamma}{5184}\notag\\
B_1&=&\f{A_1}{10\,368} + \(\f{283}{373\,248} - \f{5\g}{5184}\) \f{\L}{E_\text{Ry}}  \notag\\
B_2&=&\f{A_2}{10\,368} + \(\f{629}{1\,244\,160} - \f{\g}{1296}\) \(\f{\L}{E_\text{Ry}}\)^2\,.
\end{eqnarray}
It follows that $\de$ may be put in a form identical to equations \eqref{delta_summary} and \eqref{generic_delta}.

\bibliography{EQM_bib}
\end{document}